\documentclass[fleqn,usenatbib]{mnras}

\usepackage{newtxtext,newtxmath,bm}
\usepackage{lipsum}  

\usepackage[T1]{fontenc}

\DeclareRobustCommand{\VAN}[3]{#2}
\let\VANthebibliography\thebibliography
\def\thebibliography{\DeclareRobustCommand{\VAN}[3]{##3}\VANthebibliography}

\usepackage{graphicx}	
\usepackage{amsmath}	
\usepackage{orcidlink}
\usepackage{times}
\usepackage{xspace}
\usepackage{comment}
\usepackage{listings}
\usepackage{nth}

\usepackage{siunitx}
\DeclareSIUnit\kpc{kpc}
\DeclareSIUnit\Mpc{Mpc}
\DeclareSIUnit\Gpc{Gpc}
\DeclareSIUnit\Gyr{Gyr}

\newcommand{\codename}[1]{\textcolor{black}{\sc #1}\xspace}        
\newcommand{\simulationname}[1]{\textcolor{black}{\sc #1}\xspace}  

\newcommand{\swift}{\codename{Swift}}
\newcommand{\class}{\codename{Class}}

\newcommand{\flamingo}{\simulationname{Flamingo}}

\newcommand{\lcdm}{$\Lambda$CDM\xspace}

\title[Analytic baryonic matter power spectrum suppression]{An analytic redshift-independent formulation of baryonic effects on the matter power spectrum}

\author[M. Schaller \& J. Schaye]{
Matthieu
Schaller\,\textsuperscript{\orcidlink{0000-0002-2395-4902}}$^{\,1,2}$\thanks{E-mail:
  \url{mschaller@lorentz.leidenuniv.nl}} \&
Joop Schaye\textsuperscript{\orcidlink{0000-0002-0668-5560}}$^{2}$

\\
$^{1}$Lorentz Institute for Theoretical Physics, Leiden University, PO Box 9506, NL-2300 RA Leiden, The Netherlands\\
$^{2}$Leiden Observatory, Leiden University, PO Box 9513, NL-2300 RA Leiden, The Netherlands
}

\date{Accepted XXX. Received YYY; in original form ZZZ}

\pubyear{\the\year{}}

\begin{document}
\label{firstpage}
\pagerange{\pageref{firstpage}--\pageref{lastpage}}
\maketitle

\begin{abstract}
Baryonic effects created by feedback processes associated with galaxy formation
are an important, poorly constrained systematic effect for models of large-scale
structure as probed by weak gravitational lensing. Upcoming surveys require fast
methods to predict and marginalize over the potential impact of baryons on the
total matter power spectrum. Here we use the \flamingo cosmological
hydrodynamical simulations to test a recent proposal to approximate the matter
power spectrum as the sum of the linear matter power spectrum and a constant
multiple, $A_{\rm mod}$, of the difference between the linear and non-linear
gravity-only power spectra. We show that replacing this constant multiple with a
one-parameter family of sigmoid functions of the wavenumber $k$ allows to us
match the predictions of simulations with different feedback strengths for $z
\leq 1, k < 3~h\cdot{\rm Mpc}^{-1}$, and the different cosmological models in
the \flamingo suite. The baryonic response predicted by \flamingo models that
use jet-like AGN feedback instead of the fiducial thermally-driven AGN feedback
can also be reproduced, but at the cost of increasing the number of parameters
in the sigmoid function from one to three. The assumption that $A_{\rm mod}$
depends only on $k$ breaks down for decaying dark matter models, highlighting
the need for more advanced baryon response models when studying cosmological
models that deviate strongly from \lcdm.
\end{abstract}
\begin{keywords}
large-scale structure of Universe --  cosmology: theory -- methods: numerical
\end{keywords}



\section{Introduction}
\label{sec:introduction}

In the modern era of cosmology surveys, many probes (e.g. galaxy clustering,
cosmic shear, CMB lensing, etc.) focus on detailed measurements of the
distribution of matter in the Universe at multiple epochs and across different
length scales. With the so-called ``Stage IV'' probes now starting to collect
data, the onus is on the theorists to make accurate predictions that match the
expected quality of the data. One of the particularly challenging aspects is the
exploitation of information deep in the non-linear regime where perturbation
theory is not sufficient anymore.

In addition to the non-linear gravitational evolution of the large-scale
structure, predicting the effect galaxy formation processes and the feedback
they induce on the matter density field is especially challenging. The modelling
of these effects can be done via full hydrodynamical simulations of galaxy
formation \citep[e.g.][]{COWLS, BAHAMAS, Delgado2023, Schaye2023, Pakmor2023,
  Bigwood2025} but the immense computing resources that they require prevent
their direct use in survey analysis pipelines. The community has thus turned
towards semi-analytic models, often based on halo models
\cite[e.g.][]{Semboloni2013, Mead2015, Mead2020, Debackere2020}, or so-called
baryonification models \citep[e.g.][]{Schneider2015, Arico2021,
  Ferreira2024}. As such models can still be too slow to be used in cosmology
parameter searches, they are often used to train emulators which are themselves
fast enough \citep[e.g.][]{Giri2021, Arico2021}. With larger suites of
hydrodynamical simulations starting to emerge, training such emulators directly
on the output of simulations \citep[e.g.][]{Schaller2025} is a tempting prospect
for forthcoming survey analysis. Whilst these techniques can provide very
accurate matches to simulations (typically one per cent accuracy up to
$k\approx10~h\cdot{\rm Mpc}^{-1}$), it can be desirable to have simpler,
analytic, approximations for the baryon effects on the matter power
spectrum. Quick evaluation of the baryon response would extend its applicability
to more areas, such as the construction of covariance matrices. \\

As part of their analysis, \cite{Amon2022} decomposed the total matter power
spectrum as follows:
\begin{equation}
    P_{\rm m}\left(k,z\right) = P_{\rm m}^{\rm L}\left(k,z\right) + A_{\rm mod}\left(k,z\right) \left[P_{\rm m}^{\rm NL}\left(k,z\right) - P_{\rm m}^{\rm L}\left(k,z\right)\right],
    \label{eq:response:PS}
\end{equation}
with the function $A_{\rm mod}\left(k,z\right)$ capturing the effect
baryons have on the total matter power spectrum and $P_{\rm m}^{\rm
  L}\left(k,z\right)$, $P_{\rm m}^{\rm NL}\left(k,z\right)$
corresponding to the linear and non-linear matter power spectra. The
case $A_{\rm mod}\left(k,z\right) = 1$ would lead to a model where the
baryons have no effect on the matter density field.

\cite{Amon2022} then made two key assumptions about the function $A_{\rm
  mod}\left(k,z\right)$ they had just introduced:
\begin{enumerate}
\item $A_{\rm mod}(k,z)$ does \emph{not} depend on redshift.
\item $A_{\rm mod}(k,z)$ does \emph{not} depend on the background cosmology.
\end{enumerate}
With these two assumptions, the total, baryon-corrected, matter power
spectrum can thus be very easily evaluated as it only depends on the
linear and non-linear (i.e. gravity-only) matter power spectra for
which many rapid estimation techniques exist.

In their study, \cite{Amon2022} used a constant for the function $A_{\rm
  mod}(k,z)$ whose value was obtained by combining lensing and CMB data and
requesting a consistent cosmology fit. In their follow-up study,
\cite{Preston2023}  used a binned version of $A_{\rm mod}(k,z)$
where the value of $A_{\rm mod}$ in five different $k$-bins was obtained using a 
similar technique for different dataset combinations. We extend this here by
using an analytic function for $A_{\rm mod}$ rather than discrete bins.

The first of the remaining ingredients in eq.~\ref{eq:response:PS}, the linear
matter power spectrum, $P_{\rm m}^{\rm L}\left(k,z\right)$, is traditionally
obtained using Boltzmann solvers \citep[e.g.][]{CAMB, CLASS}. The second
ingredient, the non-linear power spectrum in a dark matter only (DMO) universe,
$P_{\rm m}^{\rm NL}\left(k,z\right)$, is often expressed in terms of the
non-linear boost to the linear power spectrum
\begin{equation}
    \beta^{\rm NL}(k,z) \equiv \frac{P_{\rm m}^{\rm NL} (k,z) }{P_{\rm m}^{\rm L} (k,z)}.
    \label{eq:response:beta}
\end{equation}
Over the last two decades, various approaches have been proposed to compute this
boost. The ``halo model'' formalism \citep{Seljak2000, halofit, Asgari2023}
proposes an analytic formalism with free parameters usually calibrated to the
results of $N$-body simulations \citep[e.g.][]{Takashi2012, Mead2016}. More
recently, interpolation between a large suite of DMO runs using various
emulation techniques has started to compete with this approach
\citep[e.g.][]{Heitmann2016, Lawrence2017, DeRose2019, Euclid2019, Bocquet2020,
  Angulo2021, StoreyFisher2024, Zhao2025}.

With these two ingredients, the response of the matter power spectrum due to
baryons can be expressed as
\begin{align}
  R(k,z) &\equiv 
         \frac{P_{\rm m} (k,z) }{P_{\rm m}^{\rm NL} (k,z)} \label{eq:response:response}\\
         &= \beta^{\rm NL}(k,z)^{-1} + A_{\rm mod}\left(k\right)
  \left[1 -  \beta^{\rm NL}(k,z)^{-1}\right]. \label{eq:response:response2}
\end{align}
Under the two assumptions listed above, $\beta^{\rm NL}(k,z)$ captures all the
cosmology and redshift dependence, and $A_{\rm mod}(k)$ encodes the specifics of
the galaxy formation model.

In this paper, we use the \flamingo suite of simulations \citep{Schaye2023,
  Kugel2023} to obtain an analytic functional form for $A_{\rm mod}(k)$ that
reproduces the data extracted from the runs using different strengths of stellar
and thermal AGN feedback using a single free parameter. Expanding the study to
models with jet-like AGN feedback, we find that the same function fits the
results well at the cost of two extra free parameters. We then use our fits to
verify whether the two key assumptions listed above hold for our simulations. In
this process, we also identify the range of redshifts where the approximation
holds, allowing the controlled use of our analytic fit within the analysis
pipeline of modern surveys.

The \flamingo simulations have been shown to reproduce a series of observables
of the galaxy and cluster population \citep{Schaye2023, Braspenning2023,
  Braspenning2025, McCarthy2023, McCarthy2024, Kumar2025}. As such, they are a
great test-bed to measure the effect of baryons on the matter power
spectrum. Furthermore, the use of variations of the base model, where the
observables have been systematically shifted in proportion to the estimated
errors on the galaxy stellar mass function and cluster gas fractions
\citep{Kugel2023}, allows for a direct connection between the baryonic response
-- here in the form of $A_{\rm mod}(k)$ -- and the observables the simulations
were calibrated to. Additionally, the simulations themselves have already been
used to investigate the role of baryons on the so-called S8 tension
\citep{McCarthy2023, McCarthy2024, Elbers2024, Schaller2025}. \\

This paper is organised as follows. In Sec.~\ref{sec:simulations} we present the
simulations used in this study. We then introduce our approximation in
Sec.~\ref{ssec:correction} to \ref{ssec:analytic_response} before testing the
key assumptions in Sec.~\ref{ssec:evolution} and \ref{ssec:cosmology}. We
generalise the model in Sec.~\ref{ssec:jet} and summarize our findings in
Sec.~\ref{sec:conclusions}.

\section{The FLAMINGO simulations}
\label{sec:simulations}

In this section, we provide a brief summary of the key components of the
\flamingo simulations used in this study. The simulations and the strategy used
to calibrate their free parameters are described in \cite{Schaye2023}
and \cite{Kugel2023}, respectively. ~\\

The simulations were performed using the open-source \swift simulation code
\citep{SWIFT} In particular, neutrinos are evolved using the $\delta f$-method
of \cite{Elbers2021} and the gas is evolved using the SPHENIX \citep{Borrow2022}
flavour of Smoothed Particle Hydrodynamics.

The simulations include subgrid prescriptions for radiative cooling following
\cite{Ploeckinger2020}, an entropy floor at high densities and star formation
using the method of \cite{Schaye2008}, the chemical enrichment model of
\cite{Wiersma2009}, and feedback from core collapse supernova using kinetic
winds of \cite{Chaikin2023}. Supermassive black holes are modeled using
ingredients from \cite{Springel2005}, \cite{Booth2009} and \cite{Bahe2022}. AGN
feedback is modeled either as thermally-driven winds \citep{Booth2009} or by the
collimated jet model of \cite{Husko2022}.

The subgrid models were calibrated using a Gaussian process emulator, trained on
a Latin hypercube of simulations, to predict the observables as a function of
the free parameters of the subgrid models. An MCMC search was then used in
combination with the emulator to find parameters reproducing the $z=0$ galaxy
stellar mass function as well as the gas fractions in low-redshift groups and
clusters inferred from X-ray and weak-lensing data, as detailed by
\cite{Kugel2023}. Besides generating sets of simulations parameters matching the
data, they also constructed simulations where the target data is shifted by
particular amounts with respect to observations. In particular, for the cluster
gas fractions, they created different models where the observed gas fractions
are shifted up and down compared to the results by $\pm N\sigma$, where $\sigma$
is the scatter in the data \citep[see][for the exact
  definitions]{Kugel2023}. This procedure was performed for both AGN models.

All the simulations used in this study were run in a volume of $1~\rm{Gpc}$ on
the side, which is sufficient to obtain a converged baryon response of the
matter power spectrum \citep{Schaller2025}. Apart from the runs used in
Sec.~\ref{ssec:cosmology}, the simulations adopt as values of the cosmological
parameters the maximum likelihood values from the DES year 3 data release
\citep{Abbott2022} combined with external probes, i.e. their ``3$\times$2pt +
All Ext.''  model \footnote{The parameter values are given in the first row of
Table~\ref{tab:cosmological_params}.}.

The initial conditions (ICs) were generated using the \codename{MonofonIC} code
\citep{Hahn2021, Elbers2022} using a 3-fluid formalism with a separate transfer
function for each of the species.

The total matter power spectra in the hydrodynamical and gravity-only
simulations are measured as described by \cite{Schaller2025}\footnote{The raw
matter power spectra for all the simulations of the \flamingo suite have been
made publicly available on the website of the project:
\url{https://flamingo.strw.leidenuniv.nl/}.}.

\section{Analytic formulation of the baryonic response}
\label{sec:reponse}

\subsection{The $A_{\rm mod}$ correction extracted from \flamingo}
\label{ssec:correction}

\begin{figure}
\includegraphics[width=\columnwidth]{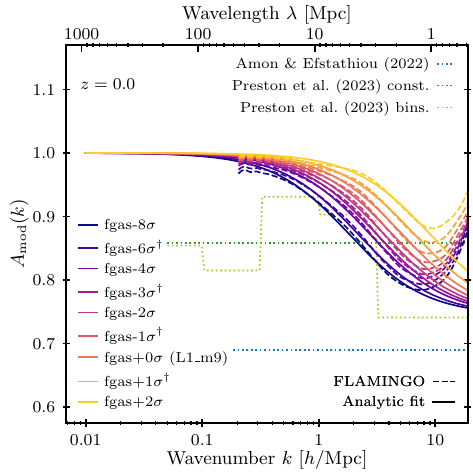}
\vspace{-0.5cm}
\caption{The function $A_{\rm mod}(k)$ at $z=0$ extracted from the \flamingo
  simulations without jet AGN calibrated to different observed gas fractions
  (dashed lines), or the emulator of \citet{Schaller2025} (indicated by a
  dagger). for intermediate gas fractions where no simulation was run. The solid
  lines in matching colours show the analytic fitting function
  (eq.~\ref{eq:response:fit}) to each \flamingo baryonic model with the
  best-fitting parameter value given in table \ref{tab:best_fit_parameters}. For
  comparison, the coloured dotted lines indicate the $A_{\rm mod}(k)$ functions
  \citet{Amon2022} and \citet{Preston2023} inferred by combining observational
  datasets.}
\label{fig:response:Amod}
\vspace{-0.3cm}
\end{figure}

We start by computing the correction function $A_{\rm mod}(k)$ at $z=0$ for the
\flamingo simulations by inverting eq.~\ref{eq:response:PS}. To compute the
linear matter power spectra for our cosmology, we use the \class \citep{CLASS}
package and add the \cite{Mead2021} halo model for the non-linear
component. Note that we do not make use of the baryonic correction their model
offers. We could alternatively have used the DMO simulations from the \flamingo
suite to obtain the non-linear matter power spectrum. Using both the linear and
non-linear power spectra from commonly used tools exemplifies how the correction
we derive here can be applied in practice without the need for additional
\flamingo data. A comparison between the \flamingo predictions and the
\cite{Mead2021} model is nevertheless shown in Appendix \ref{sec:appendix}.

The $A_{\rm mod}(k)$ corrections extracted from the simulations are shown as
dashed lines in Fig.~\ref{fig:response:Amod}. The different line colours
correspond to the corrections extracted from the various \flamingo simulations
calibrated to reproduce shifted versions of the gas fraction in clusters
inferred from X-ray and weak-lensing data. For models where no simulation data
exist, we made use of the \flamingo baryon response emulator introduced by
\cite{Schaller2025}. The \flamingo simulations are labeled by the number of
standard deviations by which the observed cluster gas fractions that the
simulations were calibrated to were shifted \citep[see][]{Kugel2023}. Rather
than considering the global shift of the dataset, it may be useful to instead
label the models by the gas fractions measured at a specific halo mass. Such a
mapping is given in Fig.~4 of \cite{Schaller2025}.

For comparison, the constant $A_{\rm mod}$ obtained by \cite{Amon2022} and
\cite{Preston2023}, as well as the binned version from \cite{Preston2023}, are
shown as dotted lines in Fig.~\ref{fig:response:Amod}. Their estimates were
constructed by combining cosmic microwave background results with weak-lensing
data and demanding a consistent cosmology fit. Note that at $k\lesssim 0.2~h
\cdot {\rm Mpc}^{-1}$, the value of $A_{\rm mod}(k)$ is not important for the
discussion that follows as the difference between the linear and non-linear
power spectra is small. This leads to large variations in $A_{\rm mod}(k)$
having only a minor impact on the total matter power spectrum. Comparing the
\flamingo curves to the ones obtained by \cite{Amon2022} and \cite{Preston2023},
we see that $A_{\rm mod}$ from the simulation is \emph{larger}, meaning that
the effect of baryons on the matter power spectrum is \emph{smaller} than their
analysis obtained. This matches the results of \cite{Schaller2025} and
\cite{McCarthy2024}. Notice also that, beyond the differences in normatlisation,
the shape of the binned $A_{\rm mod}$ by \cite{Preston2023} differs from what is
predicted by the simulations. 

\subsection{Analytic fitting functions}
\label{ssec:analytic}

\begin{table}
\centering
\caption{The values of the parameter for the best-fitting $A_{\rm mod}(k)$
  correction function (eq.~\ref{eq:response:fit}) to simulations from the
  \flamingo suite that were calibrated to different shifted versions of the
  observed gas fractions. The first column gives the simulation names used by
  \citet{Schaye2023}. The names post-fixed with a \textdagger~super-script
  indicate simulations that do not exists and whose response was obtained using
  the Gaussian-process emulator introduced by \citet{Schaller2025}.}
\label{tab:best_fit_parameters}
\begin{tabular}{l c } 
 \hline
 Simulation name & $k_{\rm mid}~[h/{\rm Mpc}]$ \\
 \hline\hline
fgas$-8\sigma$ & $1.813 \pm 0.004$ \\
fgas$-6\sigma$\textsuperscript{\textdagger} & $2.167 \pm 0.004$ \\
fgas$-4\sigma$ & $2.736 \pm 0.005$ \\
fgas$-3\sigma$\textsuperscript{\textdagger} & $3.124 \pm 0.006$ \\
fgas$-2\sigma$ & $3.614 \pm 0.008$ \\
fgas$-1\sigma$\textsuperscript{\textdagger} & $4.262 \pm 0.009$ \\
fgas$+0\sigma$~(L1$\_$m9) & $5.175 \pm 0.012$ \\
fgas$+1\sigma$\textsuperscript{\textdagger} & $6.572 \pm 0.016$ \\
fgas$+2\sigma$ & $8.968 \pm 0.026$ \\
 \hline
\end{tabular}
\end{table}

A visual inspection of the dashed curves in Fig.~\ref{fig:response:Amod}
suggests that the $A_{\rm mod}(k)$ correction extracted from the various
\flamingo simulations at $k \lesssim 10~h\cdot{\rm Mpc}^{-1}$ can be
approximated by a sigmoid curve. We choose to write:

\begin{equation}
    A_{\rm mod}(k) = A_{\rm low} + \frac{1}{2} \left(1 - A_{\rm low}\right)\left[1-\tanh\left(\frac{\log_{10}\left(k/k_{\rm mid}\right) }{\sigma_{\rm mid}} \right)\right]
    \label{eq:response:base_fit}
\end{equation}
and to fit the three free parameters to the curves extracted from the
simulations. Specifically, we use a least-square approach and use the range $k =
[0.03, 10]~h\cdot{\rm Mpc}^{-1}$ to obtain the best-fitting parameters. Note
that the choice of sigmoid guarantees that $A_{\rm mod}(k) = 1$ on
large-scales. As a result of this fitting exercise, we noticed that the values of
the parameters $\sigma_{\rm mid}$ and $A_{\rm low}$ barely varied when fitting
our function to the $A_{\rm mod}(k)$ data extracted from the runs shown in
Fig.~\ref{fig:response:Amod}. We thus decided to keep the values of these two
parameters fixed to the mean values of all the runs analysed ($\sigma_{\rm mid}
= 0.656$ and $A_{\rm low} = 0.745$). Our parametrisation of $A_{\rm mod}(k)$
thus reduces to a simple one-parameter family:

\begin{equation}
    A_{\rm mod}(k) = 0.745 + 0.1275\left[1-\tanh\left(\frac{\log_{10}\left(k/k_{\rm mid}\right) }{0.656} \right)\right].
    \label{eq:response:fit}
\end{equation}
The values of the best-fitting $k_{\rm mid}$ parameter for the individual
\flamingo simulations in which the target gas fractions for the model
calibration were varied are given in table~\ref{tab:best_fit_parameters} and the
resulting functions are shown using solid lines on
Fig.~\ref{fig:response:Amod}. As can be seen, our single-parameter sigmoids are
a good fit to the $A_{\rm mod}(k)$ extracted from the simulations up to the
wavenumber $k\approx10~h\cdot{\rm Mpc}^{-1}$ where the simulation predictions
turn over and show an increase. This is the scale where the contraction due to
baryonic cooling of the haloes matters and is also beyond the range relevant to
the current and next-generation datasets.

\subsection{The baryonic response}
\label{ssec:analytic_response}

\begin{figure}
\includegraphics[width=\columnwidth]{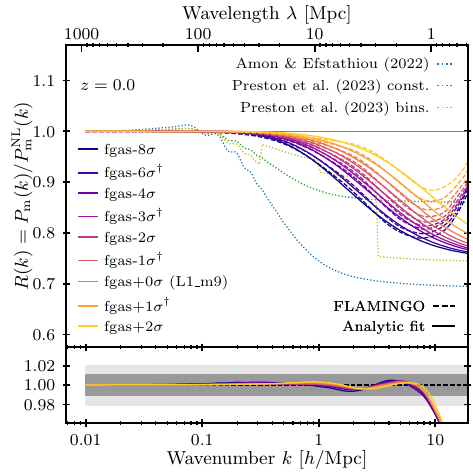}
\vspace{-0.5cm}
\caption{The baryonic response of the matter power spectrum
  (eq.~\ref{eq:response:response}) as a function of wavenumber at $z=0$ obtained
  using the $A_{\rm mod}$ correction (eq.~\ref{eq:response:PS}) and our analytic
  function (eq.~\ref{eq:response:fit}) with the parameters for each \flamingo
  model given in Table \ref{tab:best_fit_parameters} (solid lines). The dashed
  lines in matching colours correspond to the data extracted from the individual
  simulations fitted to various gas fractions in groups and clusters (or from an
  emulator, indicated by a dagger). For comparison, the coloured dotted lines
  indicate the response functions that \citet{Amon2022} and \citet{Preston2023}
  inferred by combining observational datasets. The bottom panel shows the ratio
  between our analytic model and the direct simulation prediction. The shaded
  regions correspond to fractional errors of 1 and 2 per cent respectively. For
  the different models and for all $k < 10~h \cdot {\rm Mpc}^{-1}$, the analytic
  formulation reproduces the \flamingo results to close to (or better than)
  1 per cent relative accuracy.}
\label{fig:response:response_z0}
\vspace{-0.3cm}
\end{figure}

Having constructed a simple analytic expression for $A_{\rm mod}(k)$, we turn
our attention to the baryonic response (eq.~\ref{eq:response:response}) it
generates. In Fig.~\ref{fig:response:response_z0}, we show using solid lines the
response as a function of wavenumber at $z=0$. The different colours correspond
to the various \flamingo models, as indicated in the legend. The dashed lines in
the background show the direct results of the corresponding simulations or the
results of the \cite{Schaller2025} emulator where no simulation was run. For
scales $k\lesssim 10~h\cdot{\rm Mpc}^{-1}$, our approximation is an excellent
match to the simulation (or emulator) results with all models matching the
simulations they are fitted to to better (or close to) 1 per cent accuracy. This
is of course not unexpected since we explicitly constructed the $A_{\rm mod}(k)$
sigmoid to achieve this; though the exact quantitative agreement at the level of
precision we reached with a simple one-parameter function was not necessarily to
be expected \emph{a priori}. As a point of comparison, the responses inferred by
\cite{Amon2022} and \cite{Preston2023} are showed using dotted lines. Additional
comparisons between the \flamingo results and the ones inferred from other
simulations or by analysis of observational datasets can be found in the study
by \cite{Schaller2025} (their Figs. 11 and 13, respectively).

The fact that the response inferred from \flamingo deviates from the one
obtained by \cite{Amon2022} implies that the simulations do not fit the data
they used and do not resolve the S8 tension \citep[See also][]{McCarthy2023,
  McCarthy2024, Elbers2024}. This is, at least in part, due to \cite{Amon2022}
forcing the baryonic effects to solve the tension between KiDS lensing and the
CMB and hence demanding a rather dramatic correction to the matter power
spectrum whilst the \flamingo calibration strategy was to match the observed gas
fraction in clusters. Note however, that the original tension between KiDS
lensing and the CMB seems to have dissolved in more recent analysis of the
weak-lensing data \citep{Wright2025}.

\subsection{Evolution with redshift}
\label{ssec:evolution}

\begin{figure*}
\includegraphics[width=\textwidth]{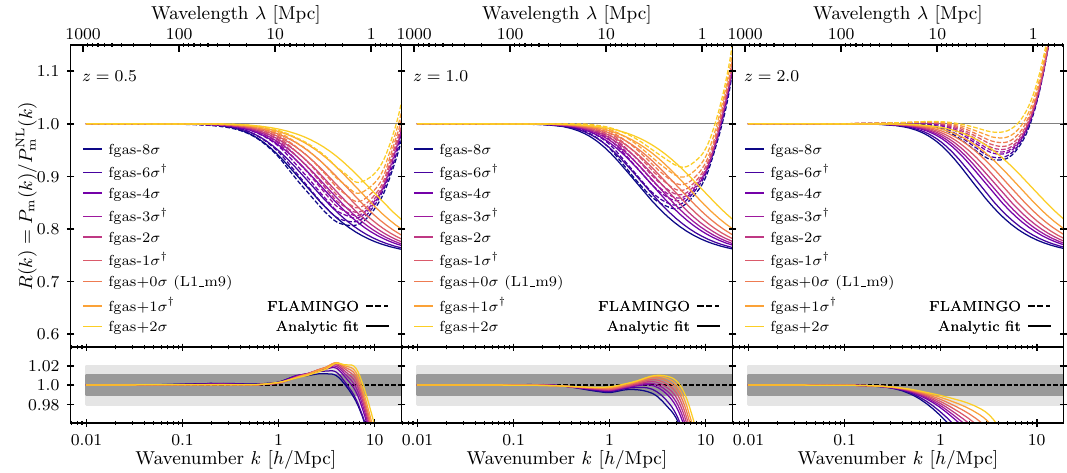}
\vspace{-0.5cm}
\caption{Same as Fig.~\ref{fig:response:response_z0} but at $z=0.5$, $1$, and
  $2$ (from left to right). Despite the fit having been performed at $z=0$, our
  analytic fit is in excellent agreement with the response extracted from the
  simulations at $z\leq 1$. This confirms the assumption that, at the level of
  precision required here and for this range of redshifts, $A_{\rm mod}(k, z)$
  does not require a redshift dependence.}
\label{fig:response:response_z}
\vspace{-0.3cm}
\end{figure*}

The first key assumption underlying the approach of writing the total matter
power spectrum in the form of eq.~\ref{eq:response:PS} is that the function
$A_{\rm mod}$ is independent of redshift. Having fitted $A_{\rm mod}(k)$ to our
$z=0$ simulation results, we now apply that fit at higher redshift. In this
process, we keep $A_{\rm mod}$ as fitted in Sec.~\ref{ssec:analytic} but compute
$P_{\rm m}^{\rm L}(k,z)$, $P_{\rm m}^{\rm NL}(k,z)$, and thus $\beta^{\rm
  NL}(k,z)$, from \class and the \cite{Mead2021} model at the redshift of
interest. We then compute the baryonic response using
eq.~\ref{eq:response:response} for our various \flamingo models and show the
results at three different redshifts in Fig.~\ref{fig:response:response_z},
where we compare it to the raw data extracted from the simulations. From left to
right, we show results at $z=0.5$, $1$, and $z=2$. The solid lines correspond to
our analytic model whilst the dashed lines show the results of the simulations
(or the emulator).

As can be seen by comparing the solid lines in the different panels, the
analytic correction displays only a small amount of evolution with redshift. The
simulation results show a more significant evolution, especially at $z > 1$. At
$z\leq 1$, the analytic $A_{\rm mod}$ model matches the simulation results at
the $2\%$ level up to scales $k = 3~h\cdot{\rm Mpc}^{-1}$ for all the models
shown here. We recall that the functional fit was only performed at $z=0$. It is
thus quite remarkable that the simple one-parameter sigmoid is sufficient to
capture the behaviour of simulations with different levels of baryonic response
over a wide range of redshifts.

We also performed the same analysis for all the intermediate redshifts where we
have data ($\Delta z = 0.05$) and measured the maximal relative error $\epsilon$
between the analytic fit and the simulation (or emulator). \\Over the range $k \leq
1~h\cdot{\rm Mpc}^{-1}$, we get:
\begin{equation}
  \epsilon < 1\% ~\mathrm{for}~  z \leq 1 \qquad \mathrm{and} \qquad \epsilon < 2\% ~\mathrm{for}~  z \leq 1.5. \nonumber
\end{equation}
Extending the range to $k \leq 3~h\cdot{\rm Mpc}^{-1}$, we find:
\begin{equation}
  \epsilon < 2\% ~\mathrm{for}~  z \leq 1 \qquad \mathrm{and} \qquad \epsilon < 5\% ~\mathrm{for}~  z \leq 1.5. \nonumber
\end{equation}
These maximal errors are measured across all the simulation variations shown in
Figs.~\ref{fig:response:Amod} to \ref{fig:response:response_z} with the value of
the single free parameter value of the fitting function given in Table
\ref{tab:best_fit_parameters}. \\

We thus conclude that our one-parameter analytic fit to $z=0$ data is
sufficiently accurate over a wide range of redshift and scales for a large range
of applications.

\subsection{Background cosmology dependence}
\label{ssec:cosmology}

\begin{table*}
\centering
\caption{The values of the cosmological parameters for the different simulations
  of the \flamingo suite used in this study. All simulations assume a flat \lcdm
  Universe including massive neutrinos with $N_{\rm eff} = 3.044$ effective
  relativistic neutrino species at high redshift and with an amount of radiation
  corresponding to $T_{\rm CMB} = 2.7255~{\rm K}$. For the simulations with
  decaying cold dark matter (DCDM), the $\Omega_{\rm cdm}$ corresponds to the
  sum of the present-day densities of decaying cold dark matter and dark
  radiation. For these models, the last column shows the dark matter decay rate,
  $\Gamma$, in units of $100~{\rm km/s/Mpc} = H_0/h$. }
\label{tab:cosmological_params}
\begin{tabular}{lcccccccccccc}
 \hline
Simulation name & $h$ & $\Omega_\text{m}$ & $\Omega_\text{cdm}$ & $\Omega_\text{b}$ & $\sum m_\nu$ & $\sigma_8$ & $10^9 A_\text{s}$ & $n_\text{s}$ & $\Gamma h/H_0$ \\
\hline
\hline
{\footnotesize D3A~(L1\_m9)}    & $0.681$ & $0.306$ & $0.256$ & $0.0486$ & $\SI{0.06}{\eV}$ & $0.807$ & $2.099$ & $0.967$ & -- \\
{\footnotesize Planck}          & $0.673$ & $0.316$ & $0.265$ & $0.0494$ & $\SI{0.06}{\eV}$ & $0.812$ & $2.101$ & $0.966$ & -- \\
{\footnotesize LS8}             & $0.682$ & $0.305$ & $0.256$ & $0.0473$ & $\SI{0.06}{\eV}$ & $0.760$ & $1.836$ & $0.965$ & -- \\
{\footnotesize PlanckNu0p24Fix} & $0.673$ & $0.316$ & $0.261$ & $0.0494$ & $\SI{0.24}{\eV}$ & $0.769$ & $2.101$ & $0.966$ & -- \\
{\footnotesize PlanckNu0p24Var} & $0.662$ & $0.328$ & $0.271$ & $0.0510$ & $\SI{0.24}{\eV}$ & $0.772$ & $2.109$ & $0.968$ & -- \\
{\footnotesize PlanckDCDM12}    & $0.673$ & $0.274$ & $0.246$ & $0.0494$ & $\SI{0.06}{\eV}$ & $0.794$ & $2.101$ & $0.966$ & $0.12$ \\
{\footnotesize PlanckDCDM24}    & $0.673$ & $0.239$ & $0.229$ & $0.0494$ & $\SI{0.06}{\eV}$ & $0.777$ & $2.101$ & $0.966$ & $0.24$ \\
\hline
\end{tabular}
\end{table*}

\begin{figure*}
\includegraphics[width=\columnwidth]{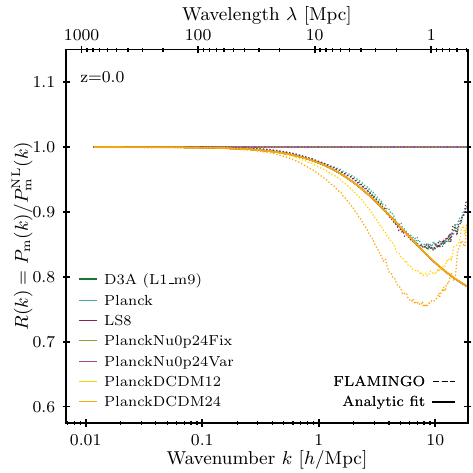}
~
\includegraphics[width=\columnwidth]{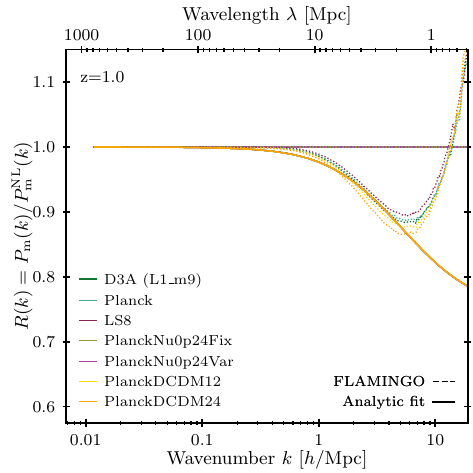}
\vspace{-0.1cm}
\caption{Same as Fig.~\ref{fig:response:response_z0} but for different
  background cosmologies in the \flamingo suite, as indicated in the legend both
  at $z=0$ (left) and $z=1$ (right). For the five cosmologies that do not
  involve decaying dark matter, the analytic formulation yields excellent
  results (the solid and dashed lines overlap), especially at $z=0$. This
  confirms the assumption that the function $A_{\rm mod}(k)$ does not require a
  dependence on the choice of background cosmology to match our simulations. We
  caution, however, that the \flamingo suite only covers a small range of
  cosmology variations. See the text for the discussion of the two models
  including decaying dark matter.}
\label{fig:response:cosmology}
\vspace{-0.3cm}
\end{figure*}

The second key assumption underlying the approach of writing the total matter
power spectrum in the form of eq.~\ref{eq:response:PS} is that the function
$A_{\rm mod}$ is independent of the chosen background cosmology. The dependence
of the total matter power spectrum on the cosmological parameters is then
entirely captured by the effects the parameters have on $P_{\rm m}^{\rm L}$ and
$P_{\rm m}^{\rm NL}$ (i.e. can be obtained from DMO simulations).

The fitting function we obtained above was constructed using \flamingo
simulations that all adopt our fiducial cosmology (D3A). We now compare the fit
to the other models that are part of the \flamingo suite. To this end, we keep
$A_{\rm mod}$ as fitted in Sec.~\ref{ssec:analytic} but compute $P_{\rm m}^{\rm
  L}(k,z)$, $P_{\rm m}^{\rm NL}(k,z)$, and thus $\beta^{\rm NL}(k,z)$, from
\class and the \cite{Mead2021} model for the different cosmologies in the
\flamingo suite whose parameter values are given in
Table~\ref{tab:cosmological_params}. We then obtain the baryonic response of the
matter power spectrum and show the results at $z=0$ using solid lines in the
left panel of Fig.~\ref{fig:response:cosmology}. We compare our analytic model
to the direct results of the simulations using dashed lines with the same
colours.

Putting aside the two models featuring decaying dark matter
(\textsc{PlanckDCDM12} and \textsc{PlanckDCDM24}), we see that the analytic
expression matches the results of the simulations extremely well. As we
performed the fit only for our fiducial cosmology, this was not guaranteed
\emph{a priori}. We see also that there is very little dependence of the
baryonic response on the cosmology. This was already noted by
\cite{Schaller2025} (their Fig. 10) and can be explained by the model proposed
by \cite{Elbers2024} who linked the changes in the response to changes in the
halo mass-concentration relation. However, we caution that the range of
cosmologies explored here does not include models that lead to large variations
in this relation and thus in the response. Two important type of parameter
variations have nevertheless been explored: those leading to scale-dependent
(changes in $n_{\rm s},\sum m_\nu$) and scale-independent (changes in
$\Omega_{\rm m}, \sigma_8$) effects on the linear power spectrum.

We also note that we do not explore models that feature significant changes in
the ratio $\Omega_{\rm b} / \Omega_{\rm m}$. We leave such explorations to
future studies that make use of a wider range of cosmological models.

It is interesting to note that, assuming a fixed $A_{\rm mod}$, the lack of a
cosmology dependence of the baryon response implies (through
eq.~\ref{eq:response:response2}) that the non-linear response $\beta^{\rm NL}$
itself is also insensitive to the choice of cosmology. This is further expanded
upon in Appendix~\ref{sec:appendix}.

On the scale of interest, $\beta^{\rm NL}$ is dominated by the one-halo term,
which displays little cosmology-dependence in halo models. It is thus
interesting to repeat the comparison at higher redshift where this term is
smaller. We show the response obtained in the simulations and using our analytic
model at $z=1$ in the right panel of Fig.~\ref{fig:response:cosmology}. For the
simulations (dashed lines), a greater level of cosmology-dependence than at
$z=0$ can be seen. This is, however, not captured by the analytic model (solid
lines). The halo model does not capture the cosmology-dependence of the one-halo
to two-halo transition. More advanced modeling of $\beta^{\rm NL}$ in the future
might reconcile the simulations and the model. We note, however, that the
differences seen here are nonetheless only at the $2\%$ level for $k <
3~h\cdot{\rm Mpc}^{-1}$.

Turning now our attention to the two models with decaying dark matter, we see
that the analytic fit to $A_{\rm mod}$ predicts the same level of response as in
the other cosmological models. This is in tension with the simulation results
(dashed lines), which predict a rather different response at $z=0$. This is in
part due to the non-linear model used to compute $P_{\rm m}^{\rm NL}$ not
reproducing the simulations in this regime (see Appendix~\ref{sec:appendix}) and
in part due to the assumption, made when constructing $A_{\rm mod}$, that the
correction does not depend on cosmology breaking down. This can be verified by
using the results of the DMO model as input $P_{\rm m}^{\rm NL}$. When doing
this, the analytic approximation also differs from the hydrodynamical simulation
results (not shown). At $z=1$ (right panel), the responses extracted from the
simulations with decaying dark matter display a behaviour closer to the other
cosmologies, and is hence captured better by our analytic model than at $z=0$.

\subsection{Other AGN feedback implementation}
\label{ssec:jet}

\begin{figure}
\includegraphics[width=\columnwidth]{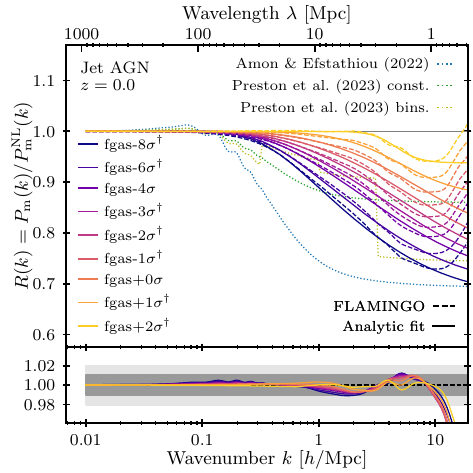}
\vspace{-0.5cm}
\caption{Same as Fig.~\ref{fig:response:response_z0} but for models
  which used the collimated jet implementation for AGN feedback instead
  of the isotropic mode (solid lines). The dashed lines in matching
  colours correspond to the data extracted from the individual
  simulations fitted to various gas fractions in groups and clusters
  (or from an emulator, indicated by a dagger). To obtain a good fit
  to the simulation results, we used the more general three-parameters
  version of $A_{\rm mod}$ (eq.~\ref{eq:response:base_fit}) with the
  values of the best-fitting parameters given in table
  \ref{tab:best_fit_parameters_jets}. The slightly more complex shape
  of the response function in the case of the collimated jet model in
  \flamingo can also be captured by our sigmoid function, albeit at
  the cost of extra free parameters.}
\label{fig:response:jets}
\vspace{-0.3cm}
\end{figure}

We have so far considered the fiducial implementation of AGN feedback in the
\flamingo suite. The set of \flamingo simulations also includes models where the
mode of injection of energy from the AGN activity was altered from
thermally-driven winds to collimated injection following the jet model of
\cite{Husko2022}. Despite these models being calibrated to the same set of
observables as the simulations using the thermal isotropic energy injection
scheme, these models display differences in the response the feedback imparts
onto the matter power spectrum \citep{Schaye2023, Schaller2025}. It is thus
interesting to extend our analysis to these models in order to provide $A_{\rm
  mod}(k)$ functions covering additional plausible scenarios.

We employ the same strategy as for the simulations using the fiducial AGN
feedback implementation (Sec.~\ref{ssec:analytic}) and fit
eq.~\ref{eq:response:base_fit} to the $A_{\rm mod}(k)$ data extracted from the
various simulations with jet AGN feedback (or from the emulator when simulations
do not exist). However, unlike in the earlier case, we find that the three free
parameters have to be varied jointly to fit the simulations. The best-fitting
parameters are provided in Table \ref{tab:best_fit_parameters_jets}.

\begin{table}
\centering
\caption{The values of the parameters for the best-fitting $A_{\rm mod}(k)$
  correction function (eq.~\ref{eq:response:base_fit}) to simulations from the
  \flamingo suite that were calibrated to different shifted versions of the
  observed gas fractions but using the collimated jets AGN feedback
  implementation. The first column gives the simulation names used by
  \citet{Schaye2023}. The names post-fixed with a super-script \textdagger
  indicate simulations that do not exist and whose response was obtained using
  the Gaussian-process emulator introduced by \citet{Schaller2025}.}
\label{tab:best_fit_parameters_jets}
\begin{tabular}{l c c c }
 \hline
 Simulation name & $k_{\rm mid}~[h/{\rm Mpc}]$ & $\sigma_{\rm mid}~[-]$ & $A_{\rm low}~[-]$\\
 \hline\hline
Jet\_fgas$-8\sigma$\textsuperscript{\textdagger}  &  $ 1.530 \pm 0.012$  &  $0.964 \pm 0.003$  &  $0.668 \pm 0.001$ \\
Jet\_fgas$-6\sigma$\textsuperscript{\textdagger}  &  $ 1.934 \pm 0.020$  &  $0.979 \pm 0.004$  &  $0.688 \pm 0.001$ \\
Jet\_fgas$-4\sigma$  &  $ 2.995 \pm 0.046$  &  $1.006 \pm 0.005$  &  $0.700 \pm 0.001$ \\
Jet\_fgas$-3\sigma$\textsuperscript{\textdagger}  &  $ 3.936 \pm 0.077$  &  $1.010 \pm 0.006$  &  $0.706 \pm 0.001$ \\
Jet\_fgas$-2\sigma$\textsuperscript{\textdagger}  &  $ 5.195 \pm 0.131$  &  $0.994 \pm 0.006$  &  $0.714 \pm 0.002$ \\
Jet\_fgas$-1\sigma$\textsuperscript{\textdagger}  &  $ 6.489 \pm 0.205$  &  $0.942 \pm 0.007$  &  $0.734 \pm 0.002$ \\
Jet\_fgas$+0\sigma$  &  $ 6.739 \pm 0.230$  &  $0.817 \pm 0.008$  &  $0.782 \pm 0.002$ \\
Jet\_fgas$+1\sigma$\textsuperscript{\textdagger}  &  $ 4.937 \pm 0.105$  &  $0.534 \pm 0.007$  &  $0.870 \pm 0.001$ \\
Jet\_fgas$+2\sigma$\textsuperscript{\textdagger}  &  $ 4.569 \pm 0.034$  &  $0.204 \pm 0.005$  &  $0.937 \pm 0.000$ \\
 \hline
\end{tabular}
\end{table}

From the best-fitting $A_{\rm mod}(k)$ curves, we construct the baryonic
response of the matter power-spectrum and show the results in
Fig.~\ref{fig:response:jets}. The dashed lines show the response extracted
directly from the simulations or emulator with the different line colours
indicating the number of sigma by which the gas fractions were shifted before
the calibration of the simulations. The solid lines in matching colours show the
best-fitting $A_{\rm mod}(k)$ function (eq.~\ref{eq:response:base_fit}) using
the best-fitting parameters of Table \ref{tab:best_fit_parameters_jets}. As can
be seen, the model is generally a good fit with a relative error (bottom panel)
of $\lesssim 1\%$ (dark grey region) up to $k\approx 10~h\cdot{\rm Mpc}^{-1}$.\\

We conclude this exercise by noting that our simple functional form is
sufficient to capture a range of baryon response behaviour, albeit at the cost
of extra free parameters.


\section{Conclusions}
\label{sec:conclusions}

In this study, we used the hydrodynamical simulations from the \flamingo project
\citep{Schaye2023, Kugel2023} to extract fitting functions for the $A_{\rm
  mod}(k,z)$ modifier that \cite{Amon2022} introduced as a simple and fast means
to model the effect of baryons on the total matter power spectrum. Here $A_{\rm
  mod}$ is the ratio between (a) the difference between the non-linear
gravity-only and linear matter power spectra and (b) the difference between the
true total non-linear matter power spectrum and the linear matter power spectrum
(eq.~\ref{eq:response:PS}). \cite{Amon2022} assumed $A_{\rm mod}(k)$ is
constant, which \cite{Preston2023} relaxed to assuming that it is independent of
redshift and cosmology

By inspecting the $A_{\rm mod}(k)$ extracted from the various simulations in the
\flamingo suite, we demonstrated that a constant is a poor approximation
(Fig.~\ref{fig:response:Amod}). We found that a sigmoid function with a single
free parameter (eq.~\ref{eq:response:fit}) is able to reproduce the $A_{\rm
  mod}(k)$ data extracted from the $z=0$ simulations calibrated to match shifted
versions of the observed gas content in clusters
(Fig.~\ref{fig:response:Amod}). We then showed that this approximation leads to a
baryonic response of the matter power spectrum matching the raw output of the
\flamingo simulations to better than $1\%$ up to wavenumbers $k \leq 10~h\cdot{\rm
  Mpc}^{-1}$ (Fig.~\ref{fig:response:response_z0}).

We then tested the key assumptions underlying the $A_{\rm mod}$ approach. We
found that the $A_{\rm mod}$ function does \emph{not} need to depend on redshift
for $z\leq1$ (Fig.~\ref{fig:response:response_z}). At $k \leq 3~h\cdot{\rm
  Mpc}^{-1}$, the maximal relative error with respect to the \flamingo results
is smaller than $2\%$ at redshifts $z\leq1$.

We also found that, within the range of cosmologies available in the \flamingo
suite, the $A_{\rm mod}$ function does \emph{not} need to depend on the choice
of cosmology (Fig.~\ref{fig:response:cosmology}). The situation is more complex
for models with decaying dark matter and, possibly, for models where the
difference with our base \lcdm cosmology increases. In particular, we note that
we have not explored models where the ratio $\Omega_{\rm b} / \Omega_{\rm m}$
varies significantly.

Finally, we explored a wider range of AGN feedback implementations and found
that our sigmoid function can also accommodate models in the \flamingo suite
that use AGN jet feedback instead of the fiducial thermally-driven AGN feedback
(Fig.~\ref{fig:response:jets}) albeit at the cost of extra parameters in the
sigmoid function. ~\\

Having verified the two key assumptions that $A_{\rm mod}(k)$ does not evolve
and does not depend on cosmology, at least for $z < 1$, CDM, and limited
variations in cosmology, we confirm that the approach proposed by
\cite{Amon2022} is valid and able to reproduce the results of complex
cosmological simulations, provided a sigmoid function is used for $A_{\rm
  mod}(k)$. Our analytic function thus provides a very efficient way, based on a
single parameter linked to the gas fraction in clusters, to obtain an excellent
estimate of the effect of baryons on the matter power spectrum at redshifts and
scales relevant to the analyses of current surveys.

\section*{Acknowledgements}

\noindent We thank the anonymous referee for their helpful comments.

This work used the DiRAC@Durham facility managed by the Institute for
Computational Cosmology on behalf of the STFC DiRAC HPC Facility
(\url{www.dirac.ac.uk}). The equipment was funded by BEIS capital funding via
STFC capital grants ST/K00042X/1, ST/P002293/1, ST/R002371/1 and ST/S002502/1,
Durham University and STFC operations grant ST/R000832/1. DiRAC is part of the
National e-Infrastructure.

\section*{Data Availability}

The raw matter power spectra used in this paper
are available on the \flamingo project's
web-page\footnote{\url{https://flamingo.strw.leidenuniv.nl/}}.

\bibliographystyle{mnras}
\bibliography{bibliography} 


\appendix
\section{Comparison of non-linear response for different cosmologies}
\label{sec:appendix}

\begin{figure}
\includegraphics[width=\columnwidth]{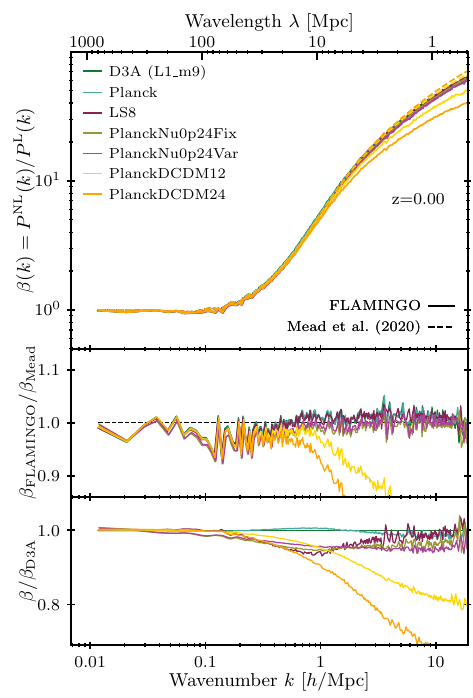}
\vspace{-0.5cm}
\caption{{\it Top:} The ratio of the non-linear and linear matter power spectra
  (eq.~\ref{eq:response:beta}) for seven different cosmologies (different
  colours) in the \flamingo suite as a function of wavenumber. The solid lines
  correspond to the results of the DMO simulations and the dashed lines in
  matching colours show the results of the \citet{Mead2021} halo model. For both
  sets of lines, the linear matter power spectrum was computed using the \class
  code \citep{CLASS}. {\it Middle:} The ratio of the non-linear response
  predicted by \flamingo and the halo model. Apart from the decaying dark matter
  models, over the entire range of scales relevant to our study, the simulations
  agree with the halo model to within a few per-cent. {\it Bottom:} The ratio of
  the non-linear response in the various cosmological models to the response in
  our fiducial cosmology (D3A) for the DMO \flamingo runs. The decaying dark
  matter models display a significantly different non-linear response from the
  other cosmological models.}
\label{fig:appendix:non_linear}
\vspace{-0.3cm}
\end{figure}

In this appendix, we compare the non-linear boost, $\beta^{\rm NL}(k,z)$,
(eq.~\ref{eq:response:beta}) extracted from DMO \flamingo simulations assuming
different cosmologies to the ones obtained using the \citep{Mead2021} halo model
as implemented in the \class code \citep{CLASS}. In both cases, we use the
\class code to obtain the linear power spectra. The boosts as a function of
wavelength extracted from the DMO \flamingo runs with different cosmologies are
shown in the top panel of Fig.~\ref{fig:appendix:non_linear} using solid
lines. The dashed lines in matching colours indicate the non-linear boosts
obtained from the halo models. The middle panel shows the ratio of the \flamingo
boost to the \citep{Mead2021} one.

Putting the two models with decaying dark matter aside, we find that the halo
model and simulations agree to within a few percent over the whole range of
scales relevant to current cosmology surveys ($k < 10~h \cdot {\rm
  Mpc}^{-1}$). The two models with decaying dark matter display a much stronger
non-linear boost in the halo model than in the \flamingo simulations. This is
not unexpected, as such cosmologies were not part of the set used to design and
test the \cite{Mead2021} halo model.

The bottom panel of Fig.~\ref{fig:appendix:non_linear}, shows the ratio of the
non-linear boost in our different cosmologies to the one extracted from the
fiducial cosmology (D3A). All the boosts here are extracted from the
simulations. As expected from the analysis in Sec.~\ref{ssec:cosmology}, we find
that the non-linear boost only has a mild dependence on cosmology.


\bsp	
\label{lastpage}
\end{document}